# Giant Current-Perpendicular-to-Plane Magnetoresistance in Multilayer-Graphene As-Grown on Nickel

S. C. Bodepudi, A. P. Singh and S. Pramanik*

Department of Electrical and Computer Engineering, University of Alberta, Edmonton, AB T6G 2V4, Canada

* **Correspondence to:** spramani@ualberta.ca

## Abstract

Strong magnetoresistance effects are often observed in ferromagnet-nonmagnet multilayers, which are exploited in state-of-the-art magnetic field sensing and data storage technologies. In this work we report a novel current-perpendicular-to-plane magnetoresistance effect in multilayer graphene as-grown on catalytic nickel surface by chemical vapor deposition. A negative magnetoresistance effect of $\sim 10^4$% has been observed, which persists even at room temperature. This effect is correlated with the shape of the $2D$ peak as well as with the occurrence of $D$ peak in the Raman spectrum of the as-grown multilayer graphene. The observed magnetoresistance is extremely high as compared to other known materials systems for similar temperature and field range, and can be qualitatively explained within the framework of "interlayer magnetoresistance" (ILMR).

[**Keywords:** Graphene, Chemical Vapor Deposition, Raman Spectroscopy, Interlayer Magnetoresistance, Current-Perpendicular-to-Plane Transport]

Artificial layered structures often exhibit strong magnetoresistance (MR) effects that are exploited in various data storage and magnetic field sensing technologies[1]. Graphite is a naturally occurring layered material in which single graphitic layers (or "graphene") are stacked on each other. Graphene, epitaxially grown on ferromagnets (such as nickel), is particularly attractive for spintronics because such systems can potentially realize perfect spin filtering[2] and giant Rashba splitting[3]. However, CPP (current-perpendicular-to-plane) MR properties of such layered graphene/ferromagnet structures are still largely underexplored. Here we consider multilayer-graphene (MLG) as-grown on nickel by chemical vapor deposition (CVD) and show that these structures exhibit large and nearly temperature-independent CPP-MR of $\sim 10^4$% for a small magnetic field of $\sim 2$ kilogauss. This MR effect is correlated with the shape of the $2D$ peak and also with the occurrence of the $D$ peak in Raman spectrum of as-grown MLG. These Raman features can be controlled by varying the CVD growth parameters. Such large negative CPP-MR, which persists even at room temperature, has hitherto not been reported in any graphitic system[4–14].

Figure 1a shows the device schematic. CVD growth of MLG is performed on 2 cm × 2 cm nickel (Ni) foils, which act as catalyst for graphene growth as well as bottom electrical contact. To ensure uniform current distribution[6], the second contact is fabricated at the center of the top MLG surface using silver epoxy. Area of the top contact is $\sim 1$ mm$^2$. As shown in Figure S1 (section I, Supplementary Information), the Ni substrate is polycrystalline with primarily (111) grains. Details of the fabrication process are provided in section I of Supplementary Information.

Figure 1b shows a FESEM image of the as-grown large-area MLG on Ni. Raman spectra taken from three representative regions of this sample are shown in the top inset of Figure 1b. The top Raman spectrum (black line) is most commonly observed, with few occurrences of the other two (blue and red). The strong $G$ peak ($\sim 1580$ cm$^{-1}$) indicates formation of hexagonal lattice of carbon atoms. The absence of disorder-induced $D$ peak ($\sim 1360$ cm$^{-1}$) indicates extremely low density of defects. We note that no $D$ peak has been found in any area of the sample. The $2D$ peak ($\sim 2700$ cm$^{-1}$) is weaker compared to the *G* peak, has a wide linewidth of ~ 50 cm$^{-1}$, and its position is slightly upshifted compared to single layer graphene[15]. These features indicate presence of multiple graphene layers[15–17], which has been independently confirmed by transferring the MLG on SiO$_2$/Si substrate[15,18] (section II, Supplementary Information). Average thickness of MLG has been found to be ~ 200 nm (Figure S2).

Unlike HOPG (Highly Oriented Pyrolytic Graphite), the *2D lineshape* of the as-grown MLG samples does not exhibit any clear "shoulder" peak at the lower frequency side of the *2D* band or any pronounced "splitting" (Figure 1b). Such absence of "shoulder" peak and splitting has been observed before in CVD-grown MLGs and turbostratic graphite [15,16,19,20]. However, such features appear after transferring the MLG on SiO$_2$/Si substrate (Figure S2).

We note that the typical penetration depth of Raman laser into graphite is ~ 50-100 nm[21–24] and hence the Raman data in Figure 1b conveys information about the graphene layers away from Ni/MLG interface. These layers are free from any crystal defect as evidenced by the absence of Raman *D* peak mentioned above. As described later, thinner MLG specimens exhibit a Raman *D* peak that originates from the defects at the Ni/MLG interface and the layers close to the interface. Thus the as-grown MLG on Ni has two distinct regions: (a) Ni/MLG interface and layers close to this interface ("defective region") and (b) layers away from Ni/MLG interface ("defect-free region").

Figure 2 shows the magnetic field (*B*) dependence of the "current-perpendicular-to-plane" (CPP) resistance ($R_{zz}$) at various temperatures (*T*). CPP-MR depends strongly on the direction of the magnetic field. For "*B* in plane" geometry ($B \perp I$ or, $\theta = \pi/2$ in Figure 1), a weak positive MR of ~ 3% or less is observed (insets of Figure 2), which gradually becomes weaker as the temperature is increased. Here magnetoresistance (MR) is defined as $[R_{zz}(11kG) - R_{zz}(0G)] \times 100\%/R_{zz}(11kG)$. However, in "*B* normal to plane" geometry ($B \parallel I$ or, $\theta = 0$ in Figure 1), resistance decreases drastically as *B* is increased and this negative MR persists over the entire temperature range of 10 − 300K.We note that the amount of resistance change is extremely large, ~ two orders of magnitude over a moderate field range of $\sim 2kG$. Semiclassical models do not predict such effects since Lorentz force on charge carriers is zero in $B \parallel I$ configuration. Multiple devices ($\sim 30$) have been tested and large resistance change (minimum by a factor of two, i.e. MR $\sim 10^2$%) has been recorded in most cases ($\sim 20$)[25].

To gain further insight into CPP charge transport we have performed temperature and magnetic field dependent current-voltage (*I-V*) measurements on the devices reported in Figure 2. Previous works have modeled CPP transport as a combination of in-plane (*xy*) charge transport in constituent layers, and phonon- or impurity-assisted interlayer (*z*) charge transfer[12,26,27]. For a disorder-free system, the latter exhibits insulator-like temperature dependence ($dR_{zz}/dT < 0$)[26,27]. In-plane resistance ($R_{in-plane}$), on the other hand, has a metal-like temperature dependence ($dR_{in-plane}/dT > 0$)[26,27]. However, recently it has been shown that in-plane resistance of CVD-grown graphene can also exhibit insulator-like temperature dependence[28]. As shown in Figure 3a, we observe an insulating behavior of the CPP resistance in our samples. In-plane resistance in our samples also exhibits insulating behavior (Supplementary Information, Section III), which is consistent with ref. [28]. However, in-plane resistance does not show any strong negative MR effect in presence of an out-of-plane magnetic field. Instead, it only shows a weak positive MR effect (Supplementary Information, Figure S3), which is consistent with previous studies[13]. In-plane transport in graphene nanoribbons can lead to large negative MR in presence of an out-of-plane magnetic field[29–31]. However this effect occurs over a field range of several Teslas and is qualitatively different than the MR effects reported in Figure 2. Thus the in-plane charge transport component is not dominant in the measured CPP resistance and the large negative MR observed in Figure 2 must originate from interlayer charge transport (along $\hat{z}$) between the graphene layers. We note that the insulating behavior of CPP resistance (or, $R_{zz}$) in Figure 2 persists both under zero field and high field conditions (Figure 3a), which excludes "magnetic field induced metal-insulator transition"[6] as the possible mechanism behind the observed large CPP-MR effect. Ref.[6] reported a positive MR in graphitic samples, which is associated with metal-insulator transition. Here, on the other hand, we observe a negative MR and our devices remain insulating over the measured magnetic field range (Figure 3a).

The current-voltage (*I-V*) characteristics are linear (Figure 3b) over a bias range of [−1mV, 1mV], which indicates that CPP-MR is independent of bias at least in ±1 mV range. The zero-field *I-V* characteristics are also linear in this bias range as shown in Figure S4 (Supplementary Information). Thus "magnetic field induced shifting of current path" appears to be an unlikely mechanism behind the observed MR effect. Further, as mentioned before such mechanism is virtually ineffective in the *B* || *I* geometry due to the absence of Lorentz force. This mechanism is strongest in the *B* ⊥ *I* geometry, but we have observed significantly weaker MR in this case (Figure 2, insets). We note that similar linear *I-V* characteristics have been reported before for *c*-axis transport in multilayer graphene stacks[12]. Resistance values measured at a higher current bias (1 mA, Figure 3a) match reasonably well with the slopes of the *I-V* curves in Figures 3b, S4. Thus it appears that the linearity of the *I-V* curves is maintained over a relatively wide current bias of 1mA. We avoid applying even larger bias in order to prevent sample damage due to large out-of-plane electric field in the CPP geometry.

To explore the role of the Ni/MLG interface and nearby graphene layers, we have studied Ni-grown MLG samples, in which the thickness of MLG is ~ 15-20 nm (Figures 3c, d). CVD growth conditions have been kept the same as before (section I, Supplementary Information). Smaller thickness of Ni substrate (~ 600 nm, e-beam evaporated) has been used in this case to obtain thinner MLG. These thinner specimens show a pronounced defective (*D*) peak in the Raman spectrum (Figure 3c), which was not present in the thicker samples (Figure 1b). This is expected since it is well known that layers close to the Ni surface do not resemble graphene-like behavior due to strong overlap between $3d$ states of Ni and $2p_z$ states of carbon[32]. For example, these interfacial layers have been known to exhibit a band gap, presence of electronic states in the band gap and absence of Dirac point[32]. Layers grown on top of these defective layers are expected to contain defects as well, which is consistent with the Raman data in Figure 3c.

We have only observed a very weak positive CPP-MR (~ 0.3%) response from this "defective region" (Figure 3d). This indicates that the large negative CPP-MR effect reported in Figure 2 does not directly originate from the "defective region" close to the Ni/MLG interface. The "defective region" also exhibits weak metallic temperature dependence of CPP resistance (Figure 3d), which is opposite of what is observed in the thicker specimens (Figure 3a). The resistance of Ni contact itself is ~ 1% of overall CPP resistance and it does not exhibit any measurable MR (Figure S5, Supplementary Information). All these evidences indicate that the transport behavior in Figures 2 and 3a, b is not limited by the "defective region" in the vicinity of Ni/MLG interface, but is governed by the "defect-free region" located away from the Ni/MLG interface. In section III of Supplementary Information we have shown that MLG/Ag paste contact resistance does not play any role either in the observed large negative CPP-MR.

To our knowledge, CPP-MR of as-grown MLG on Ni has not been studied previously and the large negative CPP-MR observed in Figure 2 has not been reported before in any graphene-based system. However, growth on Ni substrate alone is not sufficient for observation of the large negative CPP-MR effect. As described above, this effect is absent in MLG that exhibits Raman $D$ peak. The negative CPP-MR originates from the "defect-free region" away from the Ni/MLG interface and has been found to be correlated with the $2D$ lineshape in the Raman spectrum of the "defect-free region". As mentioned before, the devices characterized in Figures 1 and 2 do not exhibit any significant splitting or HOPG-like shoulder peak in the $2D$ Raman band and large negative CPP-MR is observed in these samples. We have grown MLG samples of similar thickness (~ 200 nm, on Ni) that exhibit HOPG-like shoulder peak in the Raman $2D$ band of the "defect-free region" (Figure 4a), by controlling the $CH_4:H_2$ ratio during CVD[33]. Such samples have not shown any negative CPP-MR, but showed positive CPP-MR instead, with magnetic field-dependent metal-insulator transition (Figure 4b). These features are consistent with prior CPP measurements performed on HOPG[6]. Further, we have performed CPP measurements on commercially available thick (~ 100 nm) MLG on Ni (Graphene Supermarket). The $2D$ Raman bands from the "defect-free region" of these samples exhibit pronounced splitting (Figure 4c) and these samples show no large negative CPP-MR (Figure 4d). Thus we conclude that the large negative CPP-MR originates from the graphene layers that are free of crystal defects (Raman $D$ band is absent) and that are characterized by distortion-free $2D$ Raman band. The thinner specimens (Figures 3c, d) show a distortion-free Raman 2*D* band, but it is accompanied by a Raman *D* band and no large negative CPP-MR has been observed in this case.

Negative MR in the vicinity of zero magnetic field can originate from three sources: (a) weak localization[12], (b) spin filtering and associated giant magnetoresistance (GMR) effect[2] and (c) interlayer tunneling between zero mode Landau levels[34–37]. Negative MR originating from weak localization gradually diminishes with increasing temperature due to reduction of phase coherence time at higher temperatures[12]. In our MLG/Ni samples (Figure 2) however, the negative MR is relatively insensitive to temperature. Additionally, the shape and angle dependence of MR in Figure 2 are qualitatively different than that observed in the case of weak localization and for thick samples such as ours this effect is not expected to occur[12]. Finally, the negative MR effects shown in Figure 2 are orders of magnitude stronger than a typical weak localization related effect in similar system[12]. Therefore it is unlikely that the MR response shown in Figure 2 originates from weak localization.

Very recently a novel "perfect spin filtering" effect has been proposed in MLG CVD-grown on (111) Ni[2]. Due to very high degree of spin filtering (~ 100%), such effect can give rise to very large negative MR. According to this theory, perfect spin filtering is achieved due to special energy band alignment at (111) nickel/graphene interface, which promotes transmission of *only* minority spins through the graphene[2]. Presence of multiple graphene layers (3 − 4 or more) quenches any tunneling conductance of majority spins. However in our case such spin-dependent transport cannot be used to explain the observed MR. This is because our device (Ni/MLG/Ag) is *not* a "spin valve" type device since the top electrode (Ag) is *non-magnetic*. Thus, unlike ferromagnetic contacts, this electrode is not able to differentiate between various spin orientations. Thus we believe that the observed large MR in Figure 2 is *not* due to this spin filtering effect.

Negative MR can also arise from an interlayer tunneling mechanism, which is often dubbed "interlayer magnetoresistance" or ILMR [34–37]. This effect is observed in a stack of two-dimensional (2D) massless Dirac electron systems. The interlayer coupling between these 2D layers should be sufficiently weak so that the entire system can be viewed as a stack of 2D systems instead of a bulk 3D material. In such systems out-of-plane charge transport occurs via interlayer tunneling. Such interlayer current can be tuned by a large factor by applying an out-of-plane magnetic field. The physical origin of this ILMR effect is described below.

For a stack of weakly coupled 2D massless Dirac electron system in absence of any magnetic field, electronic dispersion of each layer can be modeled by a linear *E-k* spectrum (or "Dirac cone dispersion") with Fermi level located at the Dirac points[38]. Density of states (DOS) is small at the vicinity of the Dirac point, since it depends linearly on energy measured from the Dirac point[38]. When an out-of-plane electrical bias is applied, carriers tunnel from one layer to the next. In this case small tunneling current is expected due to low DOS near the (quasi) Fermi level of each layer.

When an out-of-plane magnetic field is applied, linear *E-k* dispersion of each layer converts into a series of Landau levels, with a (zero mode) Landau level located at the Dirac point[38]. Out-of-plane charge transport will now occur via interlayer tunneling between the zero mode Landau levels. The degeneracy of the Landau levels increases with magnetic field. Thus with increasing magnetic field, degeneracy of the zero mode Landau level will increase, giving rise to larger interlayer tunneling current. This is the origin of large negative MR and is dubbed ILMR[34–37].

If the magnetic field is in-plane, it exerts Lorentz force on the electrons traveling out-of-plane and bends their trajectories. As a result, with increasing in-plane magnetic field effective interlayer tunneling distance increases, resulting in weaker tunneling probability and hence smaller interlayer tunneling current. Thus a weak positive MR is observed when the magnetic field is in-plane.

The stack of two-dimensional massless Dirac electron systems can be realized by stacking multiple graphene layers provided the interlayer coupling is sufficiently weak. Further, in order to observe the above-mentioned effect one has to ensure that out-of-plane charge transport occurs primarily via interlayer tunneling and not via some conductive defect states electrically shorting neighboring graphene layers. As described below, in our devices (Figures 1, 2) both of these conditions are fulfilled by the graphene layers in the "defect-free region" (i.e. layers away from the Ni/MLG interface).

We note that the lack of pronounced splitting (or lack of HOPG-like "shoulder peak") in the *2D* band of the Raman spectrum (Figure 1b) indicates weak interlayer coupling in the "defect-free region" of these samples. The *2D* band in the Raman spectrum of (monolayer) graphene originates due to a 2[nd] order, two-phonon, "double resonance" process, which has been analyzed extensively by various theoretical and experimental techniques[16,39–43]. Briefly, in this process first an electron-hole pair is created around K valley by a laser photon. Next, the electron is scattered by a (iTO) phonon to K' valley. Then the electron is scattered back to the K valley by another iTO phonon. This electron then recombines with a hole and emits a photon. For monolayer graphene this process leads to a single Lorentzian *2D* peak in the Raman spectrum. Now, for bilayer/trilayer graphene or graphite the single electronic dispersion curve of monolayer graphene is split into multiple branches due to interlayer coupling[38]. As a result above-mentioned double-resonance scattering processes can take place between various branches, resulting in multiple slightly-shifted Lorentzian peaks. The resultant *2D* band is a combination of all these Lorentzians and therefore loses its symmetrical shape and shows signs of splitting and distortion. In particular, for bilayer graphene four double-resonance processes can occur resulting in four Lorentzians in the vicinity of $\sim 2700$ cm$^{-1}$. The resultant *2D* band is the combination of these four components and no longer represents a Lorentzian. Similarly, for trilayer graphene fifteen different transitions are possible and the *2D* band consists of fifteen Lorentzians in the vicinity of $\sim 2700$ cm$^{-1}$ [16].

However, if interlayer coupling is weak, splitting of electronic dispersion is negligible and therefore splitting of Raman *2D* band is also negligible. We note that weak interlayer coupling and absence of splitting in the *2D* Raman band are common features of CVD-grown MLGs and have been reported by many groups in the past[15,16,19,20]. Ref.[44] employed Landau level spectroscopy to demonstrate weak interlayer coupling in graphene layers CVD-grown on Ni. The correlation between CPP-MR and *2D* lineshape as described before indicates that the observed large negative CPP-MR effect originates from the weakly coupled graphene layers.

The weak interlayer coupling and resulting (quasi) two-dimensional picture described above is valid when the nearest-neighbor interlayer coupling (or interlayer transfer energy $t_c$) is sufficiently small compared to thermal ($k_B T$) and disorder-induced ($\Gamma$) broadening. Since we have observed giant negative magnetoresistance even at the lowest temperature of $10K$ for which $\Gamma \sim 30K$[45], clearly $t_c < 30K \approx 2.6$ meV. This matches reasonably well with ref.[46], which calculated $t_c \sim 5$ $meV$ for quasi two-dimensional carriers in graphite. For HOPG samples, on the other hand, $t_c \sim 0.39$ $eV$ [14] $\gg k_B T$, $\Gamma$ and the above quasi two-dimensional picture does not hold for $T \leq 300K$. The Raman $2D$ band of HOPG is also asymmetric with a pronounced shoulder peak (Figure 4a, Figure S8). Thus no ILMR effect is expected in HOPG, which is consistent with our control experiments on HOPG-like sample (Figure 4b) and previous work[6].

As mentioned before, presence of defects is negligible in the "defect-free region" of thick MLG samples grown on Ni foil (both home-made and purchased). For MLG-on-Ni, both fcc and hcp domains can form during CVD growth, depending on the adsorption sites of the carbon atoms[47]. According to ref.[47], the grain boundaries between these domains are often "delaminated" from the substrate and MLG tends to bulge away from the substrate. These features allow matching of fcc and hcp domains by a continuous sheet of graphene, without formation of defects at the domain boundaries. In our MLG-on-Ni samples we have observed similar bulging features as shown in the Supplementary Information (Figure S6). This is presumably the reason for absence of $D$ peak in the "defect-free region" of our MLG-on-Ni samples. Thus weak interlayer coupling is preserved in the "defect-free region" of MLG-on-Ni samples described in Figures 1 and 2 and in this region CPP charge transport primarily occurs via interlayer tunneling and not via conductive defect states.

The thin MLG samples ($\sim 15 - 20$ nm) as-grown on Ni also show symmetric $2D$ Raman peak (Figure 3c). However, unlike thicker specimens, these are accompanied by defect ($D$) peak and CPP transport is fundamentally different from the thicker samples (Figure 3). In these thinner samples CPP charge transfer primarily occurs via defect states instead of any interlayer tunneling. As a result no large negative CPP-MR is observed in these thinner samples.

We have studied CPP-MR of MLG samples as-grown on Cu (Supplementary Information, section VII). Such devices have been studied by other groups in the past[12] but no large negative CPP-MR (similar to Figure 2) was reported. Samples grown on copper typically show symmetric $2D$ peak (Supplementary Information, Figure S7(a)) and therefore large negative CPP-MR is expected in these samples as well. However, the copper-grown MLG samples also exhibit significant defect peak (~1360 cm$^{-1}$, Figure S7(a)) in the Raman spectrum, which originates from the grain boundaries and also as a result of the fabrication process[12,48]. The edge states at the grain boundaries are known to be conductive[49] and can effectively short the neighboring graphene layers. In this case CPP charge transport will primarily occur via the conductive edge states instead of interlayer tunneling. This is presumably the reason for absence of large negative CPP-MR in Cu-grown samples. Transport through edge states and defects however results in a weak negative CPP-MR at low temperature due to weak localization and this effect has been observed by us (Figure S7(b)) and is consistent with literature[12].

Summarizing the above discussion, we conclude that for the samples in Figures 1, 2 the graphene layers in the "defect-free region" (i.e. away from the Ni/MLG interface) are characterized by distortion-less $2D$ band and absence of $D$ band in the Raman spectrum. Thus in this region interlayer coupling is weak and charge transfer via conductive edge states (defects) is absent. Due to weak interlayer coupling, CPP resistance is dominated by these weakly coupled layers and a large zero-field CPP resistivity is observed compared to typical HOPG samples[6]. CPP charge transfer in these layers takes place only by interlayer tunneling and results in a large negative ILMR in presence of an out-of-plane magnetic field. Thus the primary role of the Ni substrate is to realize graphene layers that are weakly coupled and are not electrically shorted with each other via conductive edge states. Our CVD growth parameters (Supplementary Information, section II) allow us to simultaneously satisfy both of these conditions.

Varying the growth parameters can give rise to strong interlayer coupling characterized by split or HOPG-like $2D$ band in Raman spectrum (Figures 4a, c). Similarly in case of commercially available HOPG, interlayer coupling is strong as well (Supplementary Information, Figure S8). Growth on Cu leads to formation of defect states at grain boundaries (Figure S7), which effectively short neighboring graphene layers. In all of these cases no ILMR has been observed. We have also transferred the Ni-grown MLG on Au and Ni electrodes (patterned on a SiO$_2$/Si substrate) and have performed CPP-MR measurements (Supplementary Information, section IX). As mentioned before, unlike as-grown MLG these transferred samples exhibit a shoulder in the 2$D$ Raman band that arises as a result of the transfer process (Figure S2). Thus interlayer coupling is adversely affected as a result of transfer and we have not observed any large negative CPP-MR, but found weak positive CPP-MR instead (Supplementary Information, Figures S9, S10), which is consistent with prior studies[6,12,14].

According to the interlayer tunneling theory[34], interlayer conductivity $\sigma_{zz}$ (which is proportional to the inverse of measured interlayer resistance $R_{zz}$) is proportional to the out-of-plane magnetic field $|B|$ and is given by:

$$\sigma_{zz} = 2C\tau|B|\frac{t_c^2 ce^3}{\pi\hbar^3}$$

where $\tau$ is the characteristic lifetime associated with Landau level broadening (or relaxation time for in-plane scattering), $t_c$ is the interlayer transfer energy estimated to be ~ 2meV for weakly coupled graphene layers (as discussed earlier), $c$ represents interlayer spacing (~ 0.342 nm for turbostratic graphite with weakly coupled graphene layers), $e$ is the electronic charge and $\hbar$ is reduced Planck constant. It has been shown[34] that $C$ is ~$1/k_BT$ for "high temperatures" for which $k_BT > t_c, \hbar/\tau$. This condition is satisfied at almost all temperatures (above ~ 30K) considered in this work and hence $C$ can be equated to $1/k_BT$. Now since $\tau$ is expected to decrease with increasing temperature, we expect the slope of $\sigma_{zz} - |B|$ curve to decrease with increasing temperature.

To check the validity of this model, first we have fitted $1/R_{zz}$ as a function of $B$ as shown in Figure 5. A clear linear fit is observed in the vicinity of ~ 2 kG, where the negative MR occurs. Further as mentioned above, the slope of $1/R_{zz}$ vs. $B$ is expected to decrease with increasing temperature, which is also consistent with Figure 5.

However we note that this model is not valid in the low field regime ($\sim 0$ kG) where inter Landau level mixing plays a dominant role[35,36]. This model also does not hold at high magnetic fields ($> \sim 3$kG) where additional level splitting (due to effects such as spin-orbit interactions) may become important. We note that a strong Rashba type splitting has been reported in graphene/Ni composites[3], which can become further pronounced in presence of high magnetic field. Such additional level splittings will clearly affect the high field CPP-MR. Such effects are not captured by the interlayer tunneling model described above. Additionally, strain in graphene layers can result in strong out-of-plane pseudo-magnetic field ($\sim 300$T), which mimics the externally applied field and gives rise to pseudo Landau levels[50]. Stacking misorientations in graphene layers can result in strong in-plane pseudo-magnetic field[51]. These effects may play a role in quantitative understanding of the observed MR effect.

The observed angle dependence of the CPP-MR response (Figure 2) is also consistent with the interlayer tunneling theory. As described before, in $B \perp I$ configuration, classical Lorentz force bends carrier trajectory to the direction parallel to the MLG plane, which reduces interlayer tunneling probability and results in a positive MR. Such bending becomes less pronounced at higher temperatures due to thermal fluctuations. As a result, the positive CPP-MR effect weakens with increasing temperature as observed in Figure 2.

 In Figure 2, the shape of the *low-field* MR response (+/- 2 kG range) is strongly dependent on temperature. At low temperatures ($10 - 80$K) the low-field MR response is "smooth-topped" with gradual variation in the neighborhood of zero field, but at higher temperature ($160$K, $250$K) the low-field MR response becomes "flat topped". This indicates that at higher temperature a competing *positive* (low-field) MR effect emerges that offsets the negative (low-field) MR and makes the net (low-field) MR flat-topped. At 300K, the positive MR effect is quite dominant and we see a weak net positive MR effect in the near vicinity of the zero field. As described below, the "interlayer tunneling" model mentioned above can be used to obtain a qualitative understanding of the temperature-dependence of low-field MR curves.

According to the interlayer tunneling model, the low-field MR response results from a competition between the following two mechanisms: (a) interlayer tunneling between the zero-mode Landau levels, which gives rise to negative MR and this effect becomes dominant as the magnetic field is increased due to an increase in Landau level degeneracy and (b) a positive MR effect, which originates from inter-Landau level mixing due to thermal broadening[35]. The positive MR effect is only effective at low field range and higher temperature, where inter-Landau level spacing is comparable to thermal broadening. In presence of inter-Landau level mixing, interlayer carrier transfer is not perpendicular to the plane, which results in a positive MR. We observed such effects for temperatures 160K and 250K (Figures 2d, e), at which the positive MR effect offsets the negative MR effect in the vicinity of zero field and results in a flat-topped MR response near zero field. At 300K (Figure 2f) the positive MR exceeds the negative MR and hence we observe a small net positive MR near zero field. At 10K, 30K and 80K (Figures 2a, b, c) thermal broadening is negligible, which suppresses the positive MR effect and only negative MR is observed at low field.

Figure S11 (Supplementary Information) shows the temperature dependence of the "switching field ($B_{SW}$)" at which sharp drop in CPP resistance takes place due to the negative MR effect. In this plot $B_{SW}$ is taken to be the field value at which resistance decreases to 50% of its zero-field value and $B_{SW}$ essentially demarcates the low field region from the high field region. We note that the switching field as defined above is not necessarily same as the "critical magnetic field" required to achieve "quantum limit"[34]. As described before, the low field MR is a result of two competing processes and negative MR effect should be dominant when the inter Landau level mixing is not significant (but not necessarily zero). In order to reduce inter Landau level mixing, spacing between neighboring Landau levels should be made comparable (or larger) than thermal broadening of the Landau levels. Inter Landau level spacing increases with magnetic field[38] and thermal broadening increases with temperature. Thus at higher temperatures, negative MR will manifest at higher magnetic field. Thus $B_{sw}$ is expected to show an increasing trend with temperature. This is consistent with Figure S11.

In conclusion, we have observed a novel giant negative CPP-MR ($> 10^4$%) in MLG as-grown on Ni, which persists even at room temperature. The negative MR manifests in thick multilayer samples in which the Raman $2D$ peak does not exhibit any significant splitting (or HOPG-like distortion) and the $D$ peak is absent. Such features have been realized by controlling the parameters of the CVD growth process. The observed data is qualitatively consistent with the "interlayer magnetoresistance" (ILMR) mechanism in which interlayer charge transfer occurs between the zero mode Landau levels of weakly coupled graphene layers. Due to large MR value and its persistence at room temperature, this effect is expected to have commercial implications and encourage further research on MLG physics and MLG growth mechanisms on ferromagnetic substrates.

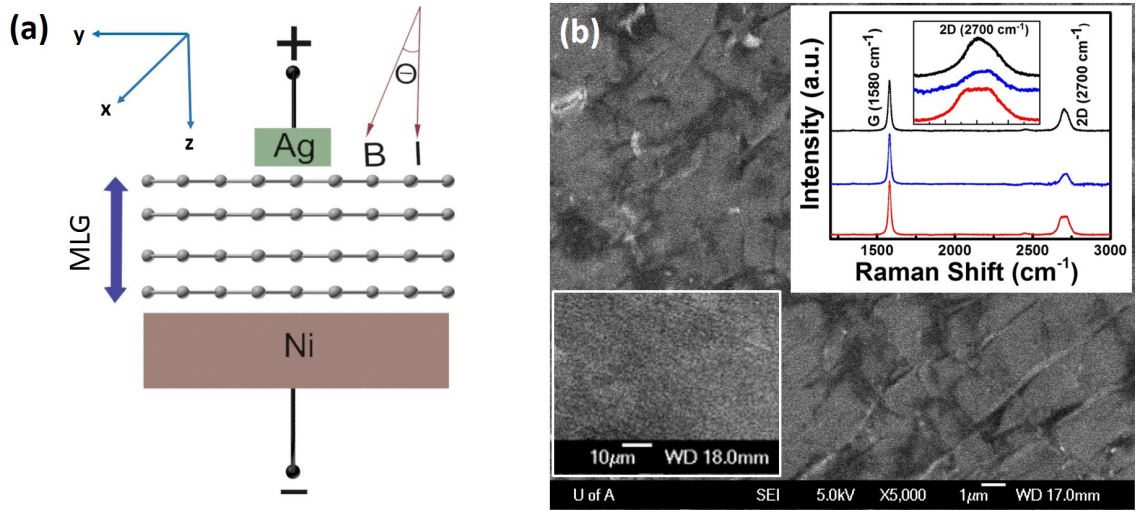

**Figure 1. Device Schematic and Raman Characterization.** (*a*) Measurement geometry. The MLG (*xy* plane) is grown on the bottom Ni substrate by CVD. CPP resistance $R_{zz}$ is measured between Ag and Ni contacts. (*b*) Field Emission Scanning Electron Microscopic (FESEM) image of as-grown MLG on Ni. Bottom inset shows the bare Ni surface. The Raman spectra (top inset) have been taken from three representative areas of the as-grown sample. The *2D* bands do not show HOPG-like "shoulder" peak or any significant splitting. The *D* peak is not present anywhere. The top Raman spectrum (black line) is most commonly observed, with few occurrences of the other two (blue and red).

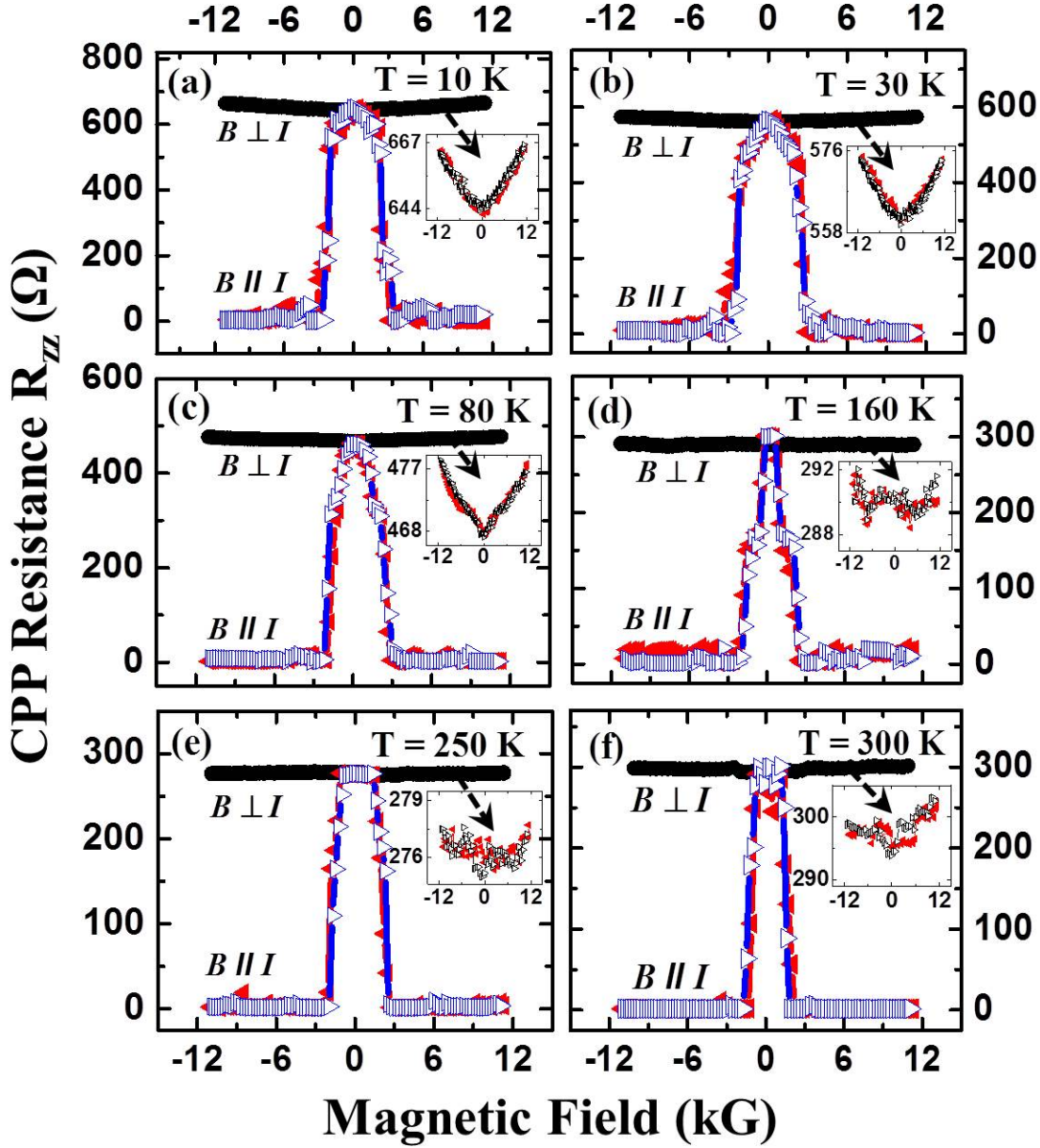

**Figure 2. Magnetic Field Dependence of CPP Resistance $R_{zz}$ in the Temperature Range 10 – 300 K for MLG/Ni Samples.** For each temperature two orientations of the magnetic field have been considered: (*i*) $B \parallel I$ ($\theta = 0$) and (*ii*) $B \perp I$ ($\theta = \pi/2$). Giant negative magnetoresistance is observed in $B \parallel I$ configuration whereas in $B \perp I$ geometry weak positive magnetoresistance is recorded (insets), which becomes less pronounced at higher temperatures. Both scan directions are shown in each plot, indicating absence of any hysteresis. Each resistance value is the average of 50 readings. For $B \parallel I$, minimum resistance (occurring at high fields) $\sim 2 - 4$ $\Omega$ as shown later in Figure 3a.

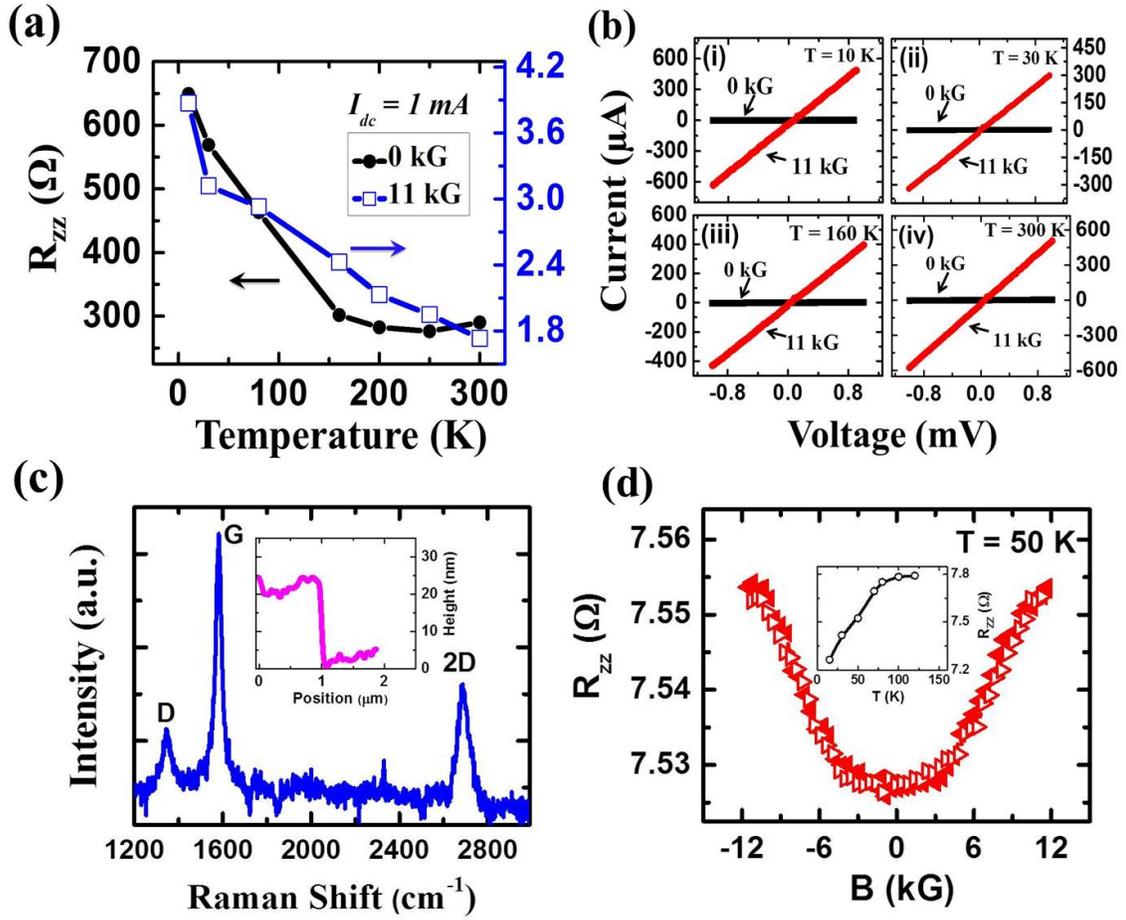

**Figure 3. CPP Charge Transport Characteristics of Thicker Samples and Control Experiments on Thinner Specimens.** (*a*) $R_{zz}(T)$ at zero field and high field indicate insulating behavior and absence of any field induced metal-insulator transition in thicker (~200 nm) samples. (*b*) $I - V$ characteristics of thicker samples at four different temperatures, indicating bias independence of the CPP-MR in ±1mV range. For zero magnetic field, current ~ few $\mu A$ at 1mV bias. A zoomed-in image of the zero-field $I - V$ data is shown in Supplementary Information, section IV. (*c*) Raman spectrum of thinner (~ 15 – 20 nm) MLG samples as-grown on Ni, which shows a defect (*D*) peak and the inset shows typical step-height measurement. (*d*) Weak positive CPP-MR and metallic temperature dependence of CPP resistance $R_{zz}$ (inset) of these thinner samples.

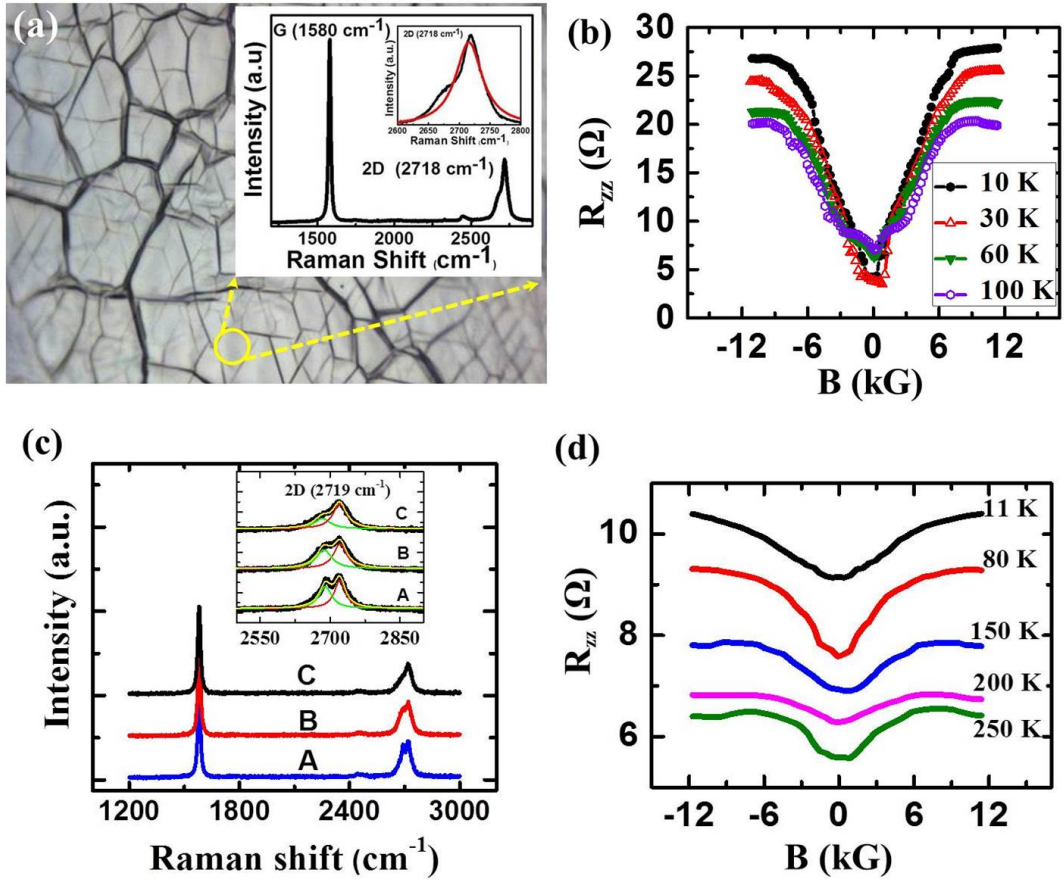

**Figure 4. Raman and CPP-MR Characterization of Control Devices (CVD-grown MLG-on-Ni – Case of Distorted *2D* Bands).** (*a*) The *2D* peak in the Raman signal shows a HOPG-like shoulder peak (*inset*). (*b*) $R_{zz}$ vs. *B* in the *B* ∥ *I* geometry. A metal-insulator transition is observed near ~1kG. (*c*), (*d*) Raman characteristics and CPP-MR of commercial MLG/Ni samples (from Graphene Supermarket). The *2D* band in the Raman spectra is clearly distorted and no negative CPP-MR has been observed.

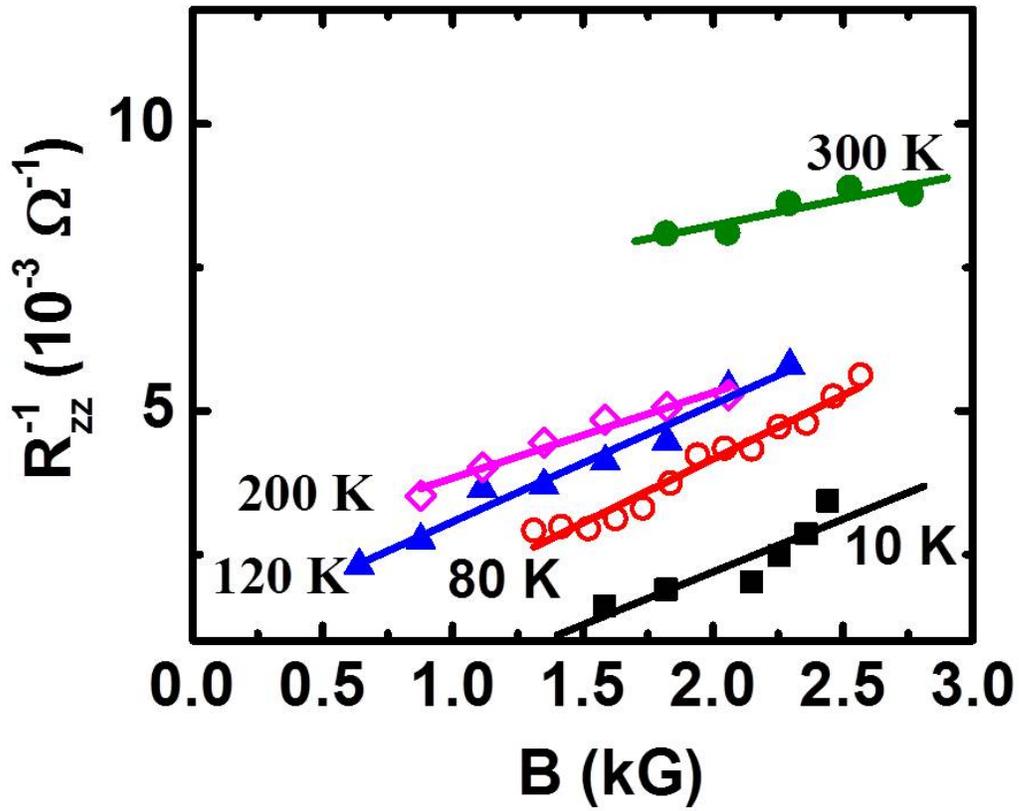

**Figure 5.** Inverse of CPP resistance ($R_{zz}^{-1}$) as a function of out-of-plane magnetic field ($B$) in the vicinity of the switching fields ($B_{sw}$). A linear trend is observed at all temperatures.

**Supplementary Information Available.**

Experimental details such as MLG growth, transfer, characterization of transferred MLG etc. are provided in the supplementary information file. This material is available free of charge via the Internet at http://pubs.acs.org.

**Acknowledgements.**

This work has been supported by TRTech (formerly TRLabs) Disruptive Technology Challenge, NSERC and SEE. We thank Mr. Rob Indoe at the National Institute for Nanotechnology (NINT) for help in sample preparation.

**References.**

(1)  Chappert, C.; Fert, A.; Dau, F. N. V. *Nature Materials* **2007**, *6*, 813–823.
(2)  Karpan, V. M.; Giovannetti, G.; Khomyakov, P. A.; Talanana, M.; Starikov, A. A.; Zwierzycki, M.; van den Brink, J.; Brocks, G.; Kelly, P. J. *Phys. Rev. Lett.* **2007**, *99*, 176602.
(3)  Dedkov, Y. S.; Fonin, M.; Rüdiger, U.; Laubschat, C. *Phys. Rev. Lett.* **2008**, *100*, 107602.
(4)  Novoselov, K. S.; Jiang, Z.; Zhang, Y.; Morozov, S. V.; Stormer, H. L.; Zeitler, U.; Maan, J. C.; Boebinger, G. S.; Kim, P.; Geim, A. K. *Science* **2007**, *315*, 1379–1379.
(5)  Dean, C. R.; Young, A. F.; Cadden-Zimansky, P.; Wang, L.; Ren, H.; Watanabe, K.; Taniguchi, T.; Kim, P.; Hone, J.; Shepard, K. L. *Nature Physics* **2011**, *7*, 693–696.
(6)  Kempa, H.; Esquinazi, P.; Kopelevich, Y. *Phys. Rev. B* **2002**, *65*, 241101.
(7)  Zhi-Ming, W.; Ding-Yu, X.; Shi-Yuan, Z.; Qing-Yu, X.; VanBael, M.; You-Wei, D. *Chinese Physics Letters* **2007**, *24*, 199–202.
(8)  Liu, Y.; Lew, W. S.; Sun, L. *Physical Chemistry Chemical Physics* **2011**, *13*, 20208.
(9)  Wu, X.; Li, X.; Song, Z.; Berger, C.; de Heer, W. A. *Phys. Rev. Lett.* **2007**, *98*, 136801.
(10)  Tikhonenko, F. V.; Kozikov, A. A.; Savchenko, A. K.; Gorbachev, R. V. *Phys. Rev. Lett.* **2009**, *103*, 226801.
(11)  Matis, B. R.; Bulat, F. A.; Friedman, A. L.; Houston, B. H.; Baldwin, J. W. *Phys. Rev. B* **2012**, *85*, 195437.
(12)  Liao, Z.-M.; Wu, H.-C.; Kumar, S.; Duesberg, G. S.; Zhou, Y.-B.; Cross, G. L. W.; Shvets, I. V.; Yu, D.-P. *Advanced Materials* **2012**, *24*, 1862–1866.
(13)  Friedman, A. L.; Tedesco, J. L.; Campbell, P. M.; Culbertson, J. C.; Aifer, E.; Perkins, F. K.; Myers-Ward, R. L.; Hite, J. K.; Eddy, C. R.; Jernigan, G. G.; Gaskill, D. K. *Nano Lett.* **2010**, *10*, 3962–3965.
(14)  Kopelevich, Y.; da Silva, R. R.; Pantoja, J. C. M.; Bratkovsky, A. M. *Physics Letters A* **2010**, *374*, 4629–4632.

(15) Reina, A.; Jia, X.; Ho, J.; Nezich, D.; Son, H.; Bulovic, V.; Dresselhaus, M. S.; Kong, J. *Nano Lett.* **2008**, *9*, 30–35.

(16) Malard, L. M.; Pimenta, M. A.; Dresselhaus, G.; Dresselhaus, M. S. *Physics Reports* **2009**, *473*, 51–87.

(17) Charlier, J.; Eklund, P.; Zhu, J.; Ferrari, A. In *Carbon Nanotubes*; Springer Berlin / Heidelberg, 2008; Vol. 111, pp. 673–709.

(18) Kim, K. S.; Zhao, Y.; Jang, H.; Lee, S. Y.; Kim, J. M.; Kim, K. S.; Ahn, J.-H.; Kim, P.; Choi, J.-Y.; Hong, B. H. *Nature* **2009**, *457*, 706–710.

(19) Faugeras, C.; Nerrière, A.; Potemski, M.; Mahmood, A.; Dujardin, E.; Berger, C.; de Heer, W. A. *Applied Physics Letters* **2008**, *92*, 011914–011914–3.

(20) Lenski, D. R.; Fuhrer, M. S. *Journal of Applied Physics* **2011**, *110*, 013720–013720–4.

(21) Ammar, M. R.; Rouzaud, J. N.; Vaudey, C. E.; Toulhoat, N.; Moncoffre, N. *Carbon* **2010**, *48*, 1244–1251.

(22) Ni, Z. hua; Wang, Y. ying; Yu, T.; Shen, Z. xiang. *arXiv:0810.2836* **2008**.

(23) Blake, P.; Hill, E. W.; Neto, A. H. C.; Novoselov, K. S.; Jiang, D.; Yang, R.; Booth, T. J.; Geim, A. K. *Applied Physics Letters* **2007**, *91*, 063124.

(24) Wang, X.; Chen, Y. P.; Nolte, D. D. *Opt. Express* **2008**, *16*, 22105–22112.

(25) Bodepudi, S.; Singh, A.; Pramanik, S. *Electronics* **2013**, *2*, 315–331.

(26) Matsubara, K.; Sugihara, K.; Tsuzuku, T. *Phys. Rev. B* **1990**, *41*, 969–974.

(27) Venugopal, G.; Jung, M.-H.; Suemitsu, M.; Kim, S.-J. *Carbon* **2011**, *49*, 2766–2772.

(28) Heo, J.; Chung, H. J.; Lee, S.-H.; Yang, H.; Seo, D. H.; Shin, J. K.; Chung, U.-I.; Seo, S.; Hwang, E. H.; Das Sarma, S. *Phys. Rev. B* **2011**, *84*, 035421.

(29) Oostinga, J. B.; Sacépé, B.; Craciun, M. F.; Morpurgo, A. F. *Phys. Rev. B* **2010**, *81*, 193408.

(30) Poumirol, J.-M.; Cresti, A.; Roche, S.; Escoffier, W.; Goiran, M.; Wang, X.; Li, X.; Dai, H.; Raquet, B. *Phys. Rev. B* **2010**, *82*, 041413.

(31) Bai, J.; Cheng, R.; Xiu, F.; Liao, L.; Wang, M.; Shailos, A.; Wang, K. L.; Huang, Y.; Duan, X. *Nat Nano* **2010**, *5*, 655–659.

(32) Wintterlin, J.; Bocquet, M.-L. *Surface Science* **2009**, *603*, 1841–1852.

(33) Chae, S. J.; Güneş, F.; Kim, K. K.; Kim, E. S.; Han, G. H.; Kim, S. M.; Shin, H.-J.; Yoon, S.-M.; Choi, J.-Y.; Park, M. H.; Yang, C. W.; Pribat, D.; Lee, Y. H. *Advanced Materials* **2009**, *21*, 2328–2333.

(34) Osada, T. *J. Phys. Soc. Jpn.* **2008**, *77*, 084711.

(35) Morinari, T.; Tohyama, T. *J. Phys. Soc. Jpn.* **2010**, *79*, 044708.

(36) Tajima, N.; Sugawara, S.; Kato, R.; Nishio, Y.; Kajita, K. *Phys. Rev. Lett.* **2009**, *102*, 176403.

(37) Tajima, N.; Sato, M.; Sugawara, S.; Kato, R.; Nishio, Y.; Kajita, K. *Phys. Rev. B* **2010**, *82*, 121420.

(38) Castro Neto, A. H.; Guinea, F.; Peres, N. M. R.; Novoselov, K. S.; Geim, A. K. *Rev. Mod. Phys.* **2009**, *81*, 109–162.

(39) Thomsen, C.; Reich, S. *Phys. Rev. Lett.* **2000**, *85*, 5214–5217.

(40) Saito, R.; Jorio, A.; Souza Filho, A. G.; Dresselhaus, G.; Dresselhaus, M. S.; Pimenta, M. A. *Phys. Rev. Lett.* **2001**, *88*, 027401.

(41) Narula, R.; Reich, S. *Phys. Rev. B* **2008**, *78*, 165422.

(42) Ferrari, A. C.; Meyer, J. C.; Scardaci, V.; Casiraghi, C.; Lazzeri, M.; Mauri, F.; Piscanec, S.; Jiang, D.; Novoselov, K. S.; Roth, S.; Geim, A. K. *Phys. Rev. Lett.* **2006**, *97*, 187401.

(43) Pimenta, M. A.; Dresselhaus, G.; Dresselhaus, M. S.; Cancado, L. G.; Jorio, A.; Saito, R. *Physical Chemistry Chemical Physics* **2007**, *9*, 1276.

(44) Luican, A.; Li, G.; Reina, A.; Kong, J.; Nair, R. R.; Novoselov, K. S.; Geim, A. K.; Andrei, E. Y. *Phys. Rev. Lett.* **2011**, *106*, 126802.

(45) Giesbers, A. J. M.; Ponomarenko, L. A.; Novoselov, K. S.; Geim, A. K.; Katsnelson, M. I.; Maan, J. C.; Zeitler, U. *Phys. Rev. B* **2009**, *80*, 201403.

(46) Haering, R. R.; Wallace, P. R. *Journal of Physics and Chemistry of Solids* **1957**, *3*, 253–274.

(47) Lahiri, J.; Lin, Y.; Bozkurt, P.; Oleynik, I. I.; Batzill, M. *Nat Nano* **2010**, *5*, 326–329.

(48) Yu, Q.; Jauregui, L. A.; Wu, W.; Colby, R.; Tian, J.; Su, Z.; Cao, H.; Liu, Z.; Pandey, D.; Wei, D.; Chung, T. F.; Peng, P.; Guisinger, N. P.; Stach, E. A.; Bao, J.; Pei, S.-S.; Chen, Y. P. *Nat Mater* **2011**, *10*, 443–449.

(49) Acik, M.; Chabal, Y. J. *Japanese Journal of Applied Physics* **2011**, *50*, 070101.

(50) Levy, N.; Burke, S. A.; Meaker, K. L.; Panlasigui, M.; Zettl, A.; Guinea, F.; Neto, A. H. C.; Crommie, M. F. *Science* **2010**, *329*, 544–547.

(51) He, W.-Y.; Su, Y.; Yang, M.; He, L. *Phys. Rev. B* **2014**, *89*, 125418.

**Supporting Information.**

**Giant Current-Perpendicular-to-Plane Magnetoresistance in Multilayer Graphene as Grown on Nickel**

S. C. Bodepudi, A. P. Singh and S. Pramanik*

Department of Electrical and Computer Engineering, University of Alberta, Edmonton, Alberta T6G 2V4, Canada

* **Email:** spramani@ualberta.ca

**Experimental Details.**

**I. Growth of Multilayer Graphene (MLG) on Ni Substrates Using Chemical Vapor Deposition (CVD).**

Multilayer graphene (MLG) has been grown on polycrystalline nickel (Ni) substrates (25μm thick, annealed, 99.5% metal basis, purchased from Alfa Aesar) in a commercial Tystar chamber using a standard chemical vapor deposition (CVD) method[1–4]. **Figure S1** shows the X-ray diffraction (XRD) spectrum of the Ni foil, indicating the presence of predominantly Ni (111) grains with a small population of Ni (100) grains. Similar polycrystalline Ni substrates have been used by various groups in the past for graphene growth[1,4].

CVD process flow consists of the following steps: (a) Ni substrate (~ 2 cm × 2 cm in size) load, (b) furnace purge, (c) temperature ramp to 1000°C and hydrogen anneal for one hour, (d) graphene growth at 1000°C for 10 minutes, (e) natural cooling, and finally (f) unloading of the sample. We have used 0.3% $CH_4$, 9.7% $H_2$, and 90% Ar during the growth of graphene. This growth process results in absence of $D$ peak and absence of significant splitting in the $2D$ peak in the Raman spectra (Figure 1b, main text). These features are correlated with the observation of large negative CPP-MR (Figure 2, main text). Higher relative concentration of carbon-containing species within the chamber leads to bulk graphite (HOPG-like) growth on the Ni surface[3,5], which exhibits HOPG-like Raman $2D$ band (Figure 4a, main text). The hydrogen anneal step removes the native oxide layer and any adsorbate from the Ni surface.

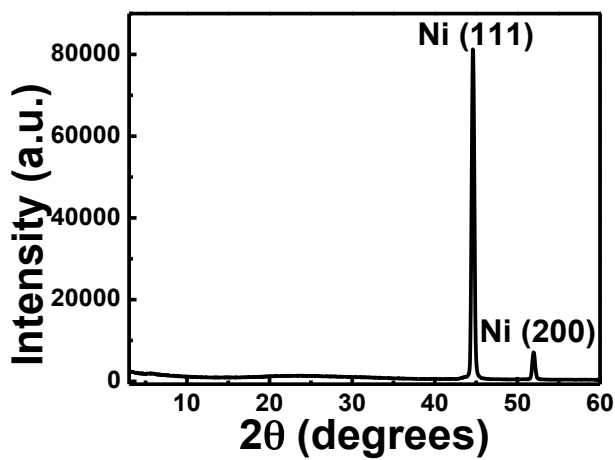

**Figure S1:** XRD spectrum of the Ni substrate used for CVD growth of MLG.

**II. Transfer of MLG on SiO$_2$/Si Substrate and Raman Characterization.**

In the main text, we have shown Raman spectra from representative regions of the as-grown MLG on Ni (Figure 1b, top inset). The MLG has also been characterized after transferring on SiO$_2$/Si substrates. Transfer of MLG has been performed as follows[2,3]. We first spin coated PMMA on the top graphene surface. Since graphene growth takes place on both surfaces of the Ni foil, we removed the back graphene layer using plasma etching and exposed the Ni substrate. (Similar steps have been followed while preparing samples for magnetotransport measurements, where the Ni substrate acts as an electrical contact). The Ni substrate was then etched away using warm ($\sim 60^{o}$C) ferric chloride (FeCl$_3$) solution. The etching process takes $\sim 4 - 5$ hours to complete. Upon completion of the etching step, MLG/PMMA composite film floats up in the FeCl$_3$ solution. MLG/PMMA film was collected and thoroughly cleaned in deionized water. Next, the cleaned MLG/PMMA film was transferred on a SiO$_2$ (300nm)/Si wafer. After overnight drying we heated the sample ($\sim 250^{o}$ C) in a vacuum oven so that the graphene layer adheres strongly to SiO$_2$ surface. Finally, top PMMA layer was dissolved in acetone, which left only MLG layer on SiO$_2$ surface.

**Figure S2 (a)** shows typical Raman characteristics of transferred MLG. All Raman spectra in this work have been acquired at room temperature using a Nicolet Almega XR Micro and Macro Raman Analysis System. Laser wavelength is set to 532nm (2.33eV). Maximum power is set to 24mW, 100% of which has been used for all studies. No evidence of laser induced heating[6] has been found in any case. Unlike as-grown MLG, the *2D* band in the Raman spectrum resembles HOPG-like *2D* band with a shoulder in the lower frequency range of the band. Such change in Raman characteristics has been observed before by other groups[1,3,7]and can be attributed to the above-mentioned thermal processing steps involved in the transfer process.

Annealing or heat treatment indeed causes "graphitization" of turbostratic graphite. For example, ref.[7] started out with a turbostratic graphite sample in which the *2D* peak is symmetric and does not show any splitting. For this sample as discussed in the main text, interaction between the basal planes is weak enough so that the splitting in the π electrons' dispersion energies does not occur resulting in a single Lorentzian in the *2D* band. However after heat treatment, the *2D* band resembles the shoulder-like shape of HOPG, indicating splitting of the π electrons' dispersion energies. Such splitting indicates presence of significant interlayer coupling as compared to the starting turbostratic sample. Presumably, the heat treatment gives rise to planar rearrangement of the carbon atoms, which stabilizes to stronger interlayer coupling (Bernal stacked) configuration after the sample is cooled. Our observation is consistent with this work because during the transfer process we have used high temperature (250°C) vacuum annealing to achieve substrate adhesion and reliable electrical measurements.

On a related note, ref.[8] observed change in the *2D* band of monolayer graphene as a function of annealing temperature. In this case, however, interlayer coupling is irrelevant since the sample consists of only one layer of graphene. In this work the change in the *2D* band has been explained by invoking the compressive stress induced by the $SiO_2$ substrate.

We did try to carry out (CPP) electrical measurements after transfer and without annealing. Unfortunately, without annealing, the electrical contact (and adhesion) between bottom electrode and transferred MLG was poor and we were unable to get reliable electrical measurements. The transferred MLG films tend to peel off the substrate in absence of annealing. Such poor contact and poor adhesion to the bottom substrate presumably results from water or air molecules trapped in a "dead space" between MLG and the bottom contact. Annealing helps to compress this "dead space" via desorption of such species and significantly improves the quality of the bottom electrical contact (after transfer) but unfortunately destroys the weak interlayer coupling. Improvement of surface adhesion upon annealing has also been observed by other groups[9,10].

There are other advantages of post-transfer annealing as well. For example, annealing is often used to desorb and decompose the chemical (primarily PMMA) residues that remain on graphene as a result of the transfer process[9–11]. Such residues cause unintentional doping of the graphene layers and results in a shift of the Fermi level from the Dirac point. Since the CPP-MR is due to interlayer tunneling between the zero mode Landau levels that are located at the Fermi levels, such shift in the Fermi levels is undesirable. Annealing is necessary to avoid this effect.

We also note that the Ni etching step involves using hot (~ 60°C) $FeCl_3$ solution for ~ 4-5 hours. Using cold/room temperature solutions does not etch Ni completely even after prolonged etching and leaves significant amount of Ni residues. Use of hot solutions for long time during etching can also adversely affect the weak interlayer coupling.

Thus such elevated temperature processes are necessary for proper transfer and electrical measurements of MLG and these processes adversely affect the interlayer coupling. Therefore as discussed in the main text we employed alternative control experiments in which such transfer steps are avoided (Figures 3c, d, 4).

**Figure S2 (b), (c)** show optical microscopic images of the transferred MLG, which indicate presence of multiple layers of graphene. In **Figure S2 (main image)** we show the step height measurement, which indicates average thickness of MLG is ~ 200 nm. Wrinkles in the graphene layer are clearly visible, which is a characteristic of CVD-grown graphene[5].

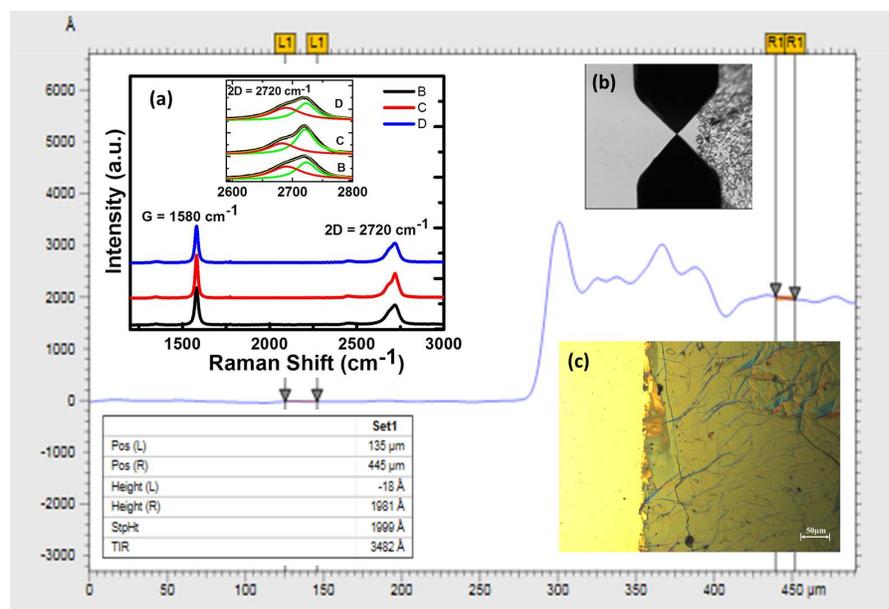

**Figure S2. Characterization of the MLG samples transferred on SiO₂/Si.** (*a*) Raman spectra, indicating HOPG-like distorted *2D* band. (*b*), (*c*) Optical images of the transferred MLG. The main image shows the step height measurement.

Transport measurements on such transferred samples have been reported in sections III and IX of this document.

### III. In-plane MR of Transferred MLG and Estimation of Ag paste/MLG Contact Resistance.

Devices for in-plane measurements have been fabricated in two different ways. In one method, MLG has been transferred on Au fingers patterned on $SiO_2$ substrate. Au wires have been attached to these fingers by using Ag paste. Thus in these devices the MLG is actually contacted by Au fingers and not directly by Ag paste. Transfer process of the MLG has been described previously in section II of this document. In the second method electrical contacts are directly made on the top surface of transferred MLG by Ag paste.

A digital image of the in-plane device and measurement geometry are shown in **Figure S3 (a)**. Distance between neighboring electrical contacts is ~ 1mm. Contact area is ~ $1mm^2$ (same as in our CPP devices). Similar in-plane device dimensions have been reported in some recent work such as ref.s [12–14].

In-plane resistance has been measured by both two-probe and four-probe methods. Devices contacted by Au fingers or Ag paste show nominally identical (in-plane) MR behavior. Since devices in which Ag paste contacts have been made directly on graphene are directly relevant for our study (since they allow estimation of Ag paste/graphene contact resistance), we present data from these devices in **Figure S3 (b)**.

Figure S3 (b) shows in-plane MR (measured in a four probe geometry, Ag-paste contacts directly made on MLG) of the transferred MLG sample. In-plane resistance exhibits insulating temperature dependence, which is consistent with ref.[15]. MR data has been taken in presence of an out-of-plane magnetic field. No strong negative MR signal has been observed. We only observe weak positive MR (~ 8% at 120K and ~ 7% at 300K). Thus the MR response reported in Figure 2 (main paper) must originate from out-of-plane carrier transport.

We have evaluated the (zero magnetic field) contact resistance between Ag paste and MLG using two different methods. First, the two-point and four-point resistances have been measured (**Figure S3 (c)**) between the inner pair of contacts (B and C) in Figure S3 (a). Contact resistance has been estimated as [two-point-resistance – four-point-resistance]/2. Two-point and four-point data have been collected by Picotest M3500 multimeter as well as Keithley 2636 Dual Channel

Source Meter. Excellent agreement has been obtained between the readings of these two equipments.

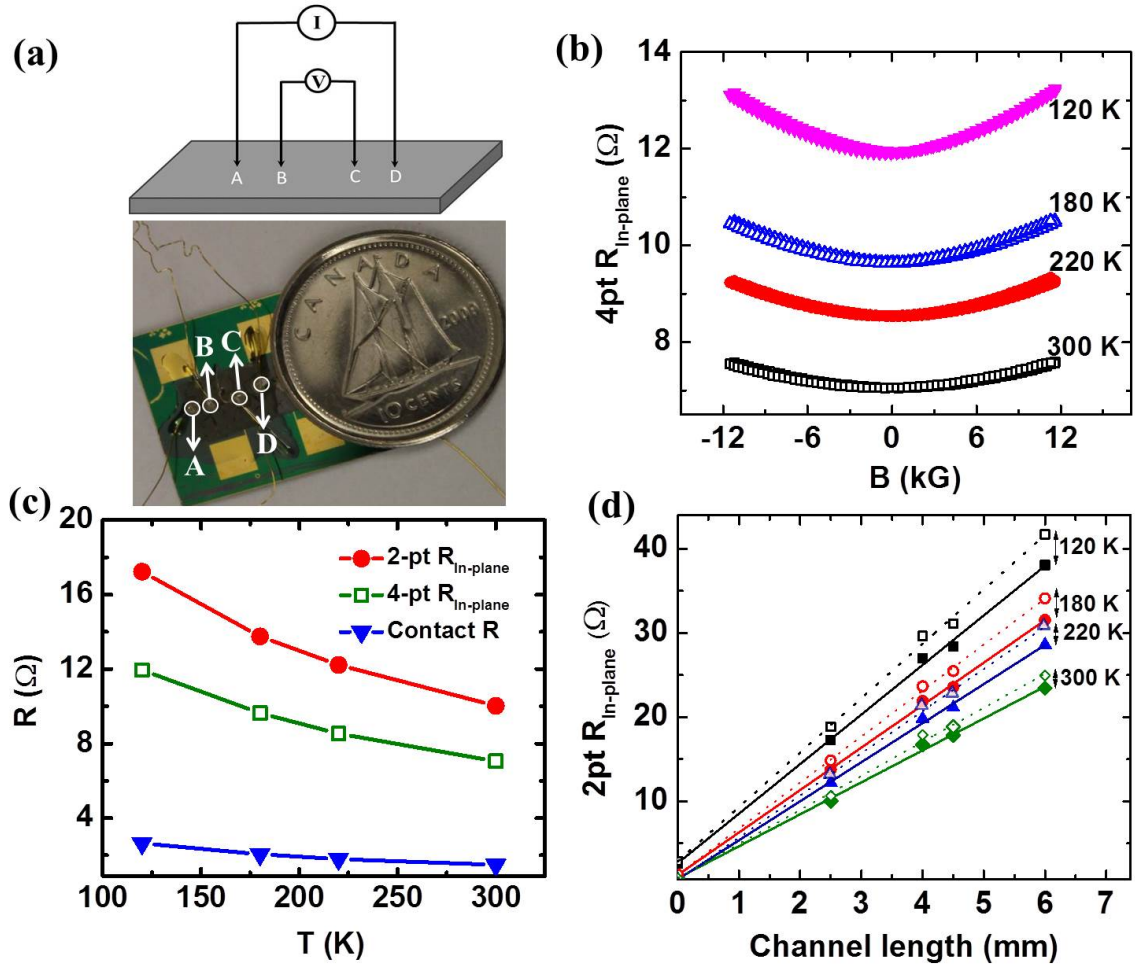

**Figure S3. In-plane Measurements.** (a) Four point measurement geometry and digital image of the device. Ag paste contacts have been made directly on MLG (contact area ~ 1mm²). Distance between neighboring contacts is ~ 1mm. Au pads have not been used in this study and they generally produce results nominally identical to those presented here. (b) In-plane four-point MR of transferred (Ni-grown) MLG. (c) Two-point and four-point in-plane resistance at zero magnetic field. Contact resistance has been estimated as half of the difference between two-point and four-point data. (d) In-plane two-point resistance vs. channel length (i.e. distance between the two contacts). Closed symbols and solid line fits correspond to zero magnetic field case whereas open symbols and dashed line fits correspond to the case when 11kG out-of-plane magnetic field is present. The linear trends extrapolated to zero channel length limit provide an estimate of the Ag paste/MLG contact resistance.

Next, we have plotted two-point resistance (zero magnetic field) as a function of the distance between contacts (or "channel length") in **Figure S3 (d)**. Two-point resistance has been found to increase linearly with increasing channel length. This linear trend, extrapolated to zero channel length limit, should give us an estimate of the contact resistance since contact resistance is independent of channel length. In Figure S3 (d) solid lines and filled symbols correspond to zero magnetic field measurements. Same current bias (1mA) has been used in all measurements (in-plane two-point and four-point in Figure S3, out-of-plane in Figures 2 and 3(a)).

Contact resistances estimated by these two methods match closely and in all cases contact resistance is $\sim 1 - 3$ Ohms. This value is two orders of magnitude smaller than the (zero field) CPP resistance reported in Figure 2. Further, the contact resistance does not depend significantly on magnetic field. In Figure S3 (d) we also show the two-point-resistance vs. channel length data taken at 11kG (open symbols, dashed lines). As expected in presence of magnetic field resistance values are slightly higher compared to the corresponding zero magnetic field values (positive in-plane MR effect). However they depend linearly on channel length and the extrapolated values in the limit of zero channel length match very well with the zero field case. Thus, Ag paste/graphene contact resistance does not play any crucial role in our reported CPP MR data.

## IV. I-V Characteristics (CPP Geometry) at Zero Magnetic Field.

**Figure S4** below shows the zoomed-in zero-field *I-V* characteristics. No non-linearity has been observed and these characteristics are very similar to the high-field *I-V* data. Thus magnetic field induced switching of current path can be ruled out as a possible cause of the observed MR effect.

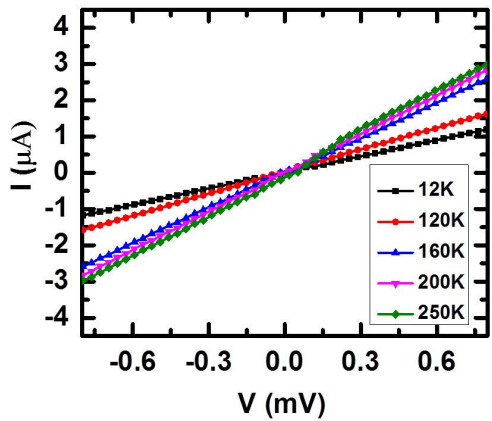

**Figure S4.** Magnified plot of the zero-field *I-V* characteristics shown in Figure 3b (main text).

## V. Magnetoresistance (MR) Measurements on Ni Substrates.

For all magnetoresistance (MR) measurements, Model 642 electromagnet power supply and Cryotronics 332 Controller (Lakeshore) have been used as electromagnet current source and temperature controller respectively. Picotest M3500 multimeter has been used to directly extract the resistance of the device. Each resistance value is the average of 50 readings. The measured resistance values have been independently validated by a Keithley Model 2636 dual-channel system source meter. A 475 DSP gauss meter has been used to record the magnetic field values. All equipments have been synchronized by LabView VI for data acquisition.

Anisotropic magnetoresistance (AMR) of the bottom Ni contact has no effect on the observed MR data due to the following reasons: (a) the typical AMR effect is ~1% (**Figure S5**), whereas we have observed ~$10^4$% MR in Figure 2, (b) the CPP resistance of the Ni substrate is ~100 times smaller than the CPP resistance of MLG/Ni composite and hence any MR of Ni substrate will be negligible in the MR of MLG/Ni composite and (c) CPP resistance of the composite exhibits a semiconducting temperature-dependence (Figure 3a, main text) and not a metallic temperature-dependence as shown below. Thus we can rule out possibility of any artifact due to AMR of Ni. Based on above considerations we can also rule out the possibility of any pinhole short between top Ag contact and bottom Ni substrate.

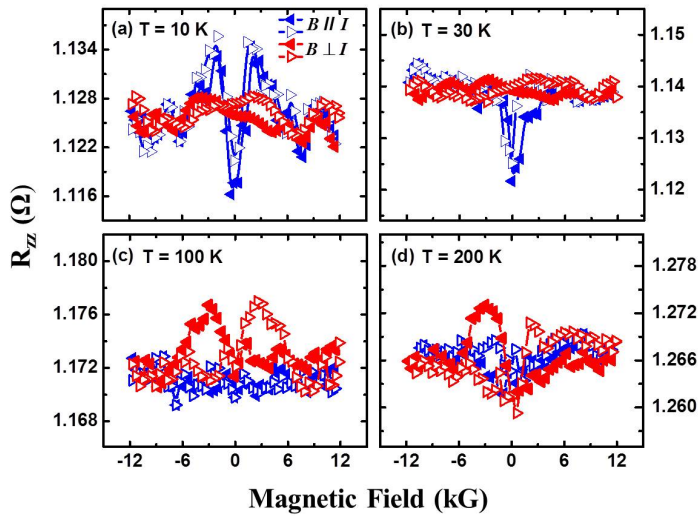

**Figure S5. Typical AMR response of Ni foil substrate.** The symbols indicate scan directions.

## VI. AFM Image of Multilayer Graphene Grown on Ni Foil.

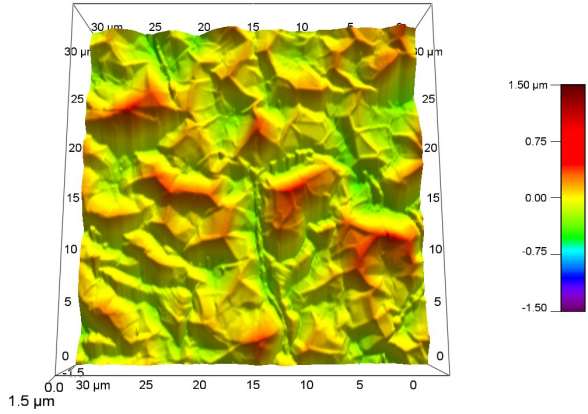

**Figure S6.** Atomic Force Microscopic (AFM) image of CVD-grown MLG-on-Ni.

We used AFM (Asylum Research, MFP-3D) under ambient conditions with a standard tetrahedral silicon tip (Olympus, OMCLAC160TS-W2) located at the end of a silicon cantilever to get surface roughness profile of CVD-grown MLG-on-Ni. Under AC mode, the typical values of force constant, resonant frequency and scan rate were 42 N/m, 300 kHz and 1 Hz respectively. The radius of curvature of the tip is < 10nm.

## VII. CPP MR of MLG CVD-Grown on Cu.

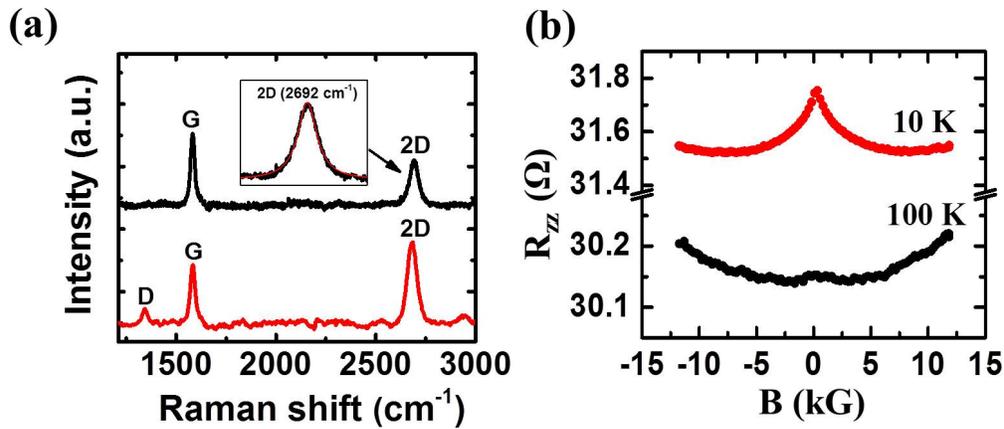

**Figure S7.** (*a*) Raman characteristics of MLG (~ 8 layer) CVD-grown on copper (ACS Materials). The bottom scan is taken from the grain boundary region, which shows a significant defect (*D*) peak. (*b*) CPP-MR of MLG/copper, showing weak localization at low field, which disappears at higher temperature. This is consistent with ref.[16], which studied MLG on Cu, prepared by a layer-by-layer transfer process.

## VIII. Raman Characterization of Highly Oriented Pyrolytic Graphite (HOPG).

**Figure S8** below shows Raman spectrum of *HOPG graphite* (SPI supplies, SPI-1 grade, 10 x 10 x 1 mm). The two characteristic peaks of HOPG occur at ~ 1580 cm⁻¹ and ~ 2718 cm⁻¹ (Figure S8) which are commonly labeled as *G* and *2D* peak respectively[17,18]. The *2D* peak for HOPG is asymmetric (inset of Figure S8), consists of two shifted Lorentzian components and the higher intensity component appears at higher frequency side.

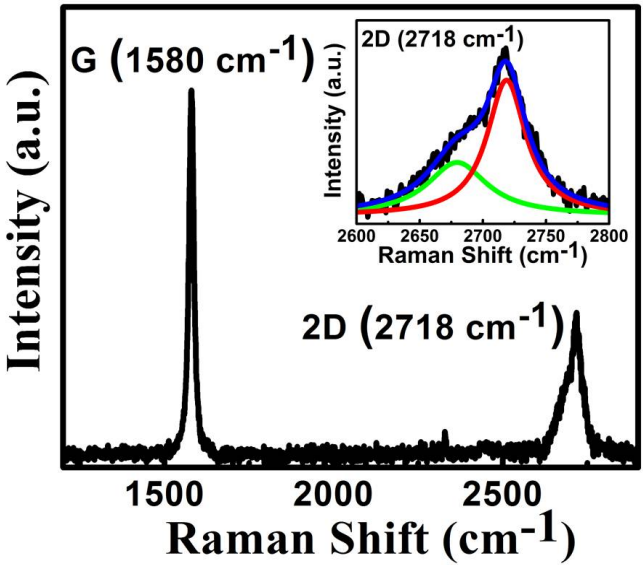

**Figure S8:** (*main image*) Raman spectrum of HOPG from SPI supplies. (*inset*) Magnified *2D* peak and the Lorentzian fits.

## IX. CPP MR of MLG Transferred on Au and Ni Electrodes.

As mentioned in the main text, the MLG samples transferred on Au and Ni electrodes do not exhibit any large negative CPP-MR, but show a weak positive CPP-MR instead. The CPP-MR responses of these transferred samples are shown below (**Figures S9, S10**). As noted before, unlike as-grown samples, the transferred MLG exhibits a shoulder in the 2*D* Raman band (Figure S2).

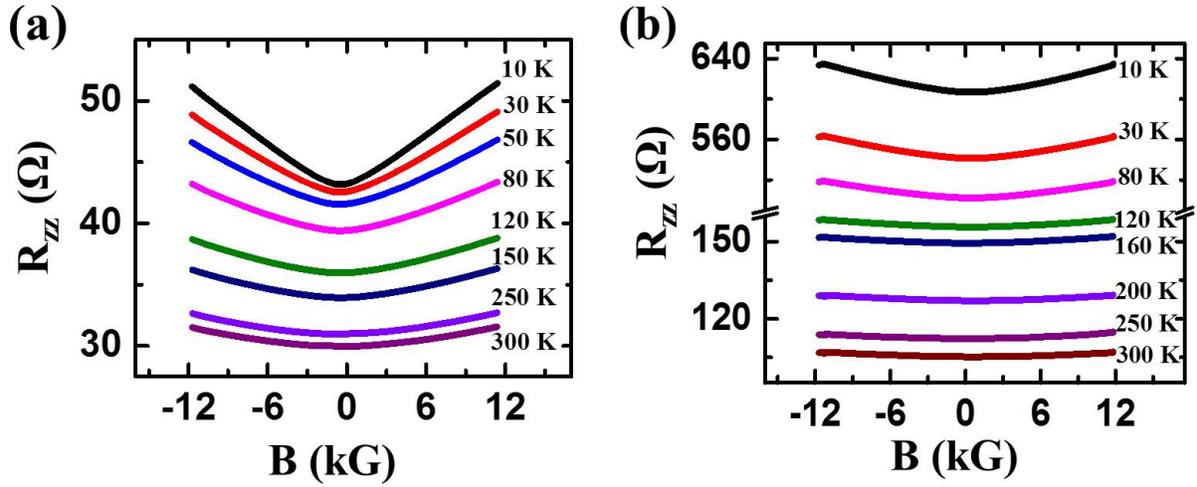

**Figure S9.** (*a*) CPP-MR of MLG (Ni-grown) transferred on Au contact and (*b*) CPP-MR of MLG (Ni-grown) transferred on Ni contact (Magnified image is shown in Figure S10 below). No weak-localization related negative MR has been observed due to presence of larger number of graphene layers. This is also consistent with ref.[16] The transferred MLG exhibits shoulder peak in Raman 2*D* band (Figure S2).

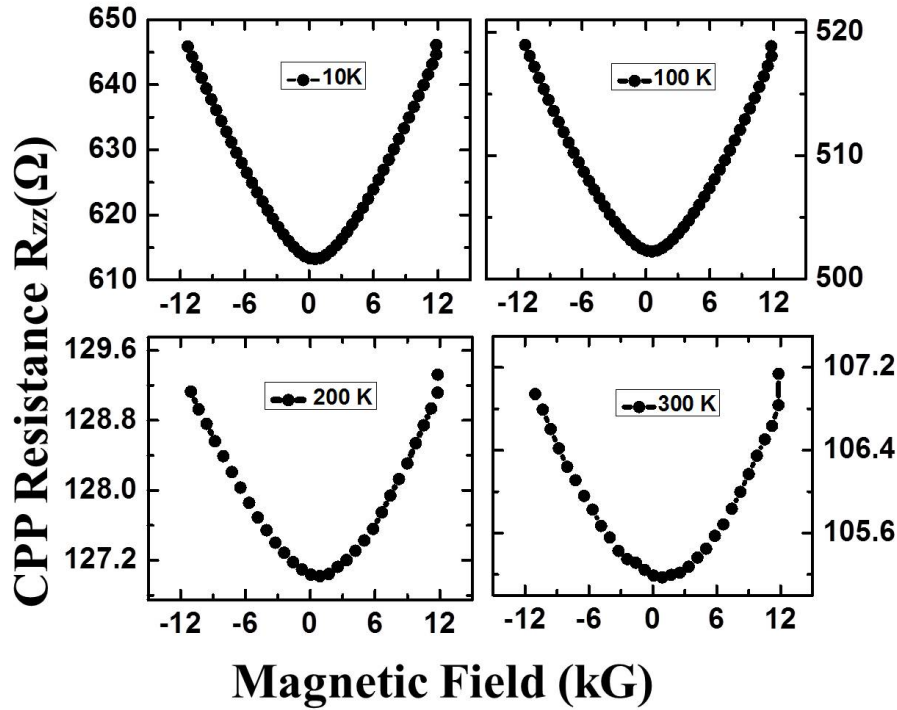

**Figure S10.** Positive CPP MR (magnified) of MLG transferred on Ni electrodes.

## X. Switching Magnetic Field as a Function of Temperature.

It is clear from Figure 2 (main text) that the sharp drop in CPP resistance takes place near a "switching magnetic field ($B_{SW}$)". **Figure S11** below shows the temperature dependence of the switching field. In this plot $B_{SW}$ is taken to be the field value at which resistance decreases to 50% of its zero-field value. A weak increasing trend has been observed which is consistent with the interlayer tunneling model since at higher temperature higher magnetic field will be needed to overcome thermal broadening.

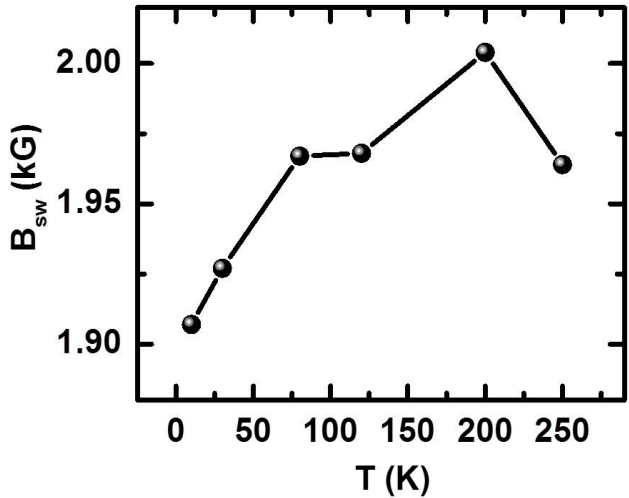

**Figure S11.** Switching magnetic field as a function of temperature.

## References

(1) Yu, Q.; Lian, J.; Siriponglert, S.; Li, H.; Chen, Y. P.; Pei, S.-S. *Applied Physics Letters* **2008**, *93*, 113103–113103–3.
(2) Reina, A.; Jia, X.; Ho, J.; Nezich, D.; Son, H.; Bulovic, V.; Dresselhaus, M. S.; Kong, J. *Nano Lett.* **2008**, *9*, 30–35.
(3) Kim, K. S.; Zhao, Y.; Jang, H.; Lee, S. Y.; Kim, J. M.; Kim, K. S.; Ahn, J.-H.; Kim, P.; Choi, J.-Y.; Hong, B. H. *Nature* **2009**, *457*, 706–710.
(4) Zhang, Y.; Gomez, L.; Ishikawa, F. N.; Madaria, A.; Ryu, K.; Wang, C.; Badmaev, A.; Zhou, C. *J. Phys. Chem. Lett.* **2010**, *1*, 3101–3107.
(5) Chae, S. J.; Güneş, F.; Kim, K. K.; Kim, E. S.; Han, G. H.; Kim, S. M.; Shin, H.-J.; Yoon, S.-M.; Choi, J.-Y.; Park, M. H.; Yang, C. W.; Pribat, D.; Lee, Y. H. *Advanced Materials* **2009**, *21*, 2328–2333.
(6) Everall, N. .; Lumsdon, J.; Christopher, D. . *Carbon* **1991**, *29*, 133–137.

(7)  Cançado, L. G.; Takai, K.; Enoki, T.; Endo, M.; Kim, Y. A.; Mizusaki, H.; Speziali, N. L.; Jorio, A.; Pimenta, M. A. *Carbon* **2008**, *46*, 272–275.

(8)  Ni, Z. hua; Wang, Y. ying; Yu, T.; Shen, Z. xiang. *arXiv:0810.2836* **2008**.

(9)  Pirkle, A.; Chan, J.; Venugopal, A.; Hinojos, D.; Magnuson, C. W.; McDonnell, S.; Colombo, L.; Vogel, E. M.; Ruoff, R. S.; Wallace, R. M. *Applied Physics Letters* **2011**, *99*, 122108.

(10) Cheng, Z.; Zhou, Q.; Wang, C.; Li, Q.; Wang, C.; Fang, Y. *Nano Lett.* **2011**, *11*, 767–771.

(11) Lin, Y.-C.; Lu, C.-C.; Yeh, C.-H.; Jin, C.; Suenaga, K.; Chiu, P.-W. *Nano Lett.* **2012**, *12*, 414–419.

(12) Wang, Y. Y.; Burke, P. J. *Applied Physics Letters* **2013**, *103*, 052103.

(13) Yang, G.; Lee, C.; Kim, J.; Ren, F.; Pearton, S. J. *Phys. Chem. Chem. Phys.* **2013**, *15*, 1798–1801.

(14) Shen, T.; Wu, W.; Yu, Q.; Richter, C. A.; Elmquist, R.; Newell, D.; Chen, Y. P. *Applied Physics Letters* **2011**, *99*, 232110.

(15) Heo, J.; Chung, H. J.; Lee, S.-H.; Yang, H.; Seo, D. H.; Shin, J. K.; Chung, U.-I.; Seo, S.; Hwang, E. H.; Das Sarma, S. *Phys. Rev. B* **2011**, *84*, 035421.

(16) Liao, Z.-M.; Wu, H.-C.; Kumar, S.; Duesberg, G. S.; Zhou, Y.-B.; Cross, G. L. W.; Shvets, I. V.; Yu, D.-P. *Advanced Materials* **2012**, *24*, 1862–1866.

(17) Malard, L. M.; Pimenta, M. A.; Dresselhaus, G.; Dresselhaus, M. S. *Physics Reports* **2009**, *473*, 51–87.

(18) Charlier, J.; Eklund, P.; Zhu, J.; Ferrari, A. In *Carbon Nanotubes*; Springer Berlin / Heidelberg, 2008; Vol. 111, pp. 673–709.